\documentclass[times,3p]{elsarticle}

\usepackage{amsmath,amssymb,latexsym}
\usepackage{graphicx}
\usepackage{enumitem}
\usepackage{url}
\usepackage{xurl}
\usepackage{xcolor}
\usepackage{hyperref}

\journal{Engineering Applications of Artificial Intelligence}
\biboptions{numbers,sort&compress}

\begin{document}

\begin{frontmatter}

\title{A Case-Bundle Operating Model for Coding Agents in OpenFOAM-Based CFD}

\author[AISI]{Ke Xiao}

\author[PKU,AISI]{Han Li\corref{cor1}}
\ead{han\_li@pku.edu.cn}

\author[AISI]{Teng Zhang}

\author[PKU]{Yangchen Xu}

\author[PKU]{Runze Mao}

\author[PKU,AISI]{Zhi X. Chen\corref{cor1}}
\ead{chenzhi@pku.edu.cn}

\cortext[cor1]{Corresponding authors}

\address[PKU]{State Key Laboratory of Turbulence and Complex Systems, College of Engineering, Peking University, Beijing 100871, China}
\address[AISI]{AI for Science Institute (AISI), Beijing 100080, China}

\begin{abstract}
General-purpose coding agents can set up computational fluid dynamics (CFD) cases, execute solvers, and manage remote jobs. Reviewable and reusable work additionally depends on persistent engineering context and evidence. We present a case-bundle operating model with two modes. Build supports agent-assisted case development under engineering review. Replay applies a reviewed case to new variants. We used this model in an OpenFOAM-7 \texttt{interFoam} study for screening injector designs. GPT-5.5 in Codex helped develop a case bundle containing the simulation configuration, geometry-processing and meshing procedures, remote-execution scripts, post-processing code, and review records. The bundle was replayed to execute and post-process 140 Stereolithography (STL) geometry variants on a remote high-performance computing system. A separate replay exercise used the Pi coding agent as the runtime with four different LLM backends. All four runs succeeded and produced verified results. Tool use and token consumption varied across runs. The results show how reviewed case bundles can support bounded, reusable automation with distinct roles for routine execution and engineering judgment.
\end{abstract}


\end{frontmatter}

\section{Introduction}
\label{sec:introduction}

In only a few years, systems based on large language models (LLMs) have developed from question-answering and code-completion tools into systems that inspect artifacts, take actions, and respond to feedback. Software engineering provides a visible example of this development. Coding agents now participate in coding, repository maintenance, issue resolution, terminal-based execution, and review workflows. Studies of public software repositories document their growing adoption and participation in commits and pull requests \citep{li_rise_2025,robbes_agentic_2026}. This shift has already changed ordinary software practice. Agents are becoming part of the working environment in which code is written, reviewed, tested, and maintained, and their effects are increasingly measured at the level of complete work cycles. Factory reports 550,000 hours of development time saved across customers and a 20\% reduction in development cycle time using its Claude-powered software agents \citep{anthropic_factory_2025}; Simplex reports reductions of 40\% in screen-design time, 70\% in screen-development time, and 17\% in internal integration-testing time after adopting Codex \citep{openai_simplex_2026}.

Expectations of comparable gains from LLM agents are now reaching computer-aided engineering (CAE).
A recent review frames LLM agents as collaborators that could plan, execute, and adapt computer-aided design (CAD) and finite-element workflows \citep{guo_llm_cae_2026}. Other studies describe agentic construction of simulation workflows \citep{elisseev_agentic_2025}, computational-mechanics pipelines spanning geometry interpretation, discretization, solver execution, and engineering assessment \citep{wilke_autonomous_computational_modeling_2026}, and coordinated engineering design systems that connect CAD, computational fluid dynamics (CFD), structural analysis, and optimization \citep{xu_engineering_ai_2025}.
Commercial CAE vendors are also integrating AI into established engineering software. Ansys has introduced Engineering Copilot and AI-assisted functions across its simulation products \citep{synopsys_ansys_2026}, while Autodesk has connected Fusion to Claude through the Model Context Protocol (MCP), allowing Claude to act on structured design data and workflows \citep{autodesk_fusion_claude_2026}.
Within this setting, CFD is an active area for exploration. OpenFOAM is a widely used open-source CFD platform \citep{jasak_openfoam_2009}, and it has become a common testbed for agentic CFD systems. Existing work has investigated retrieval-augmented case support \citep{pandey_openfoamgpt_2025}, multi-agent workflow decomposition and error correction \citep{chen_metaopenfoam_2025,yue_foam_agent_2025,yue_foam_agent2_2025},  and structured domain reasoning for end-to-end simulation \citep{fan_chatcfd_2025}. More recent systems extend these capabilities with multimodal interaction \citep{yang_swarmfoam_2026} and language-guided preprocessing, execution, and post-processing \citep{xu_cfdagent_2025}.

Most of the agentic CFD works use purpose-built agent systems. General-purpose coding-agent products offer a second path. They are already being used in data infrastructure, finance, design, legal work, marketing, and data science \citep{anthropic_claude_code_teams_2025,openai_codex_data_science_2026}. OpenFOAM practice is well suited to these agents because the solver environment is organized around artifacts they can read and modify: case directories, plain-text dictionaries, mesh scripts, run scripts, scheduler jobs, logs, and post-processing outputs. Our previous assessment found that coding agents can configure and conduct a range of CFD simulations, with geometry and mesh generation presenting the greatest difficulty \citep{xiao2026preliminary}. The present paper moves from assessment to operating practice. We ask how an engineering team can use a general-purpose coding agent to deliver reviewable and reusable CFD work.

We examine this question in OpenFOAM-based CFD practice through an case study of screening injector designs through Volume of Fluid (VOF) simulations. A reviewed OpenFOAM-7 \texttt{interFoam} case was built from Stereolithography (STL) geometry through meshing, Slurm execution, and post-processing, then replayed across 140 STL design variants. The operating model has two modes. Build is the case-development mode, where engineers stay close to the loop while the agent prepares, repairs, and documents a case. Replay is the case-reuse mode, where a reviewed case is run across approved variants with the same procedures and checkpoints. In Build, engineers inspect meshes and solver behavior, judge boundary conditions and physical assumptions, and decide which changes are meaningful, while coding agents handle file edits, scripts, command execution, and routine diagnostics. In Replay, agents can be given more autonomy for operating-condition changes, job submission, result collection, and post-processing because the actions follow established procedures and review checkpoints.

The practical unit that makes this arrangement reusable is the case bundle. By case bundle, we refer to organized OpenFOAM case directories plus the supporting files needed to reproduce, inspect, and rerun the case. This bundle preserves engineering state. It can move from Build to Replay while retaining the information needed to inspect, reject, repair, or reuse the work. Accordingly, this paper makes the following contributions to the field.

\begin{enumerate}[label=\Roman*.]
  \item We define a CFD-agent workspace around OpenFOAM case bundles, where solver dictionaries, mesh and run scripts, Slurm submission files, solver and scheduler logs, post-processing outputs, manifests, and review notes are available to the agent.
  \item We formulate a bounded operating pattern with two modes, Build and Replay, in which agents perform file edits, terminal commands, and batch procedures under explicit human checkpoints.
  \item We report the operating model in an OpenFOAM-7 \texttt{interFoam} injector-screening case study with complex STL geometry, custom preprocessing, remote working directories, Slurm jobs, and automated replay over 140 design variants with solver and post-processing completion. We further examine the portability and interaction footprint of the reviewed bundle through four model backends in Pi.
  \item We provide design implications for future CFD agent environments, emphasizing structured case directories, persistent case state, executable scripts and records, and reviewable decision points.
\end{enumerate}

The remainder of this paper is organized as follows. Section~\ref{sec:methods} describes the operating modes, case-bundle method, coding agents, CFD software, and remote execution environment used in the study. Section~\ref{sec:results} presents the injector-screening workflow and the observations derived from it. Section~\ref{sec:discussion} discusses the implications for agent-assisted CFD engineering practice. Section~\ref{sec:conclusion} summarizes the main conclusions.

\section{Methods}
\label{sec:methods}

\begin{figure}[!t]
\centering
\includegraphics[width=\textwidth]{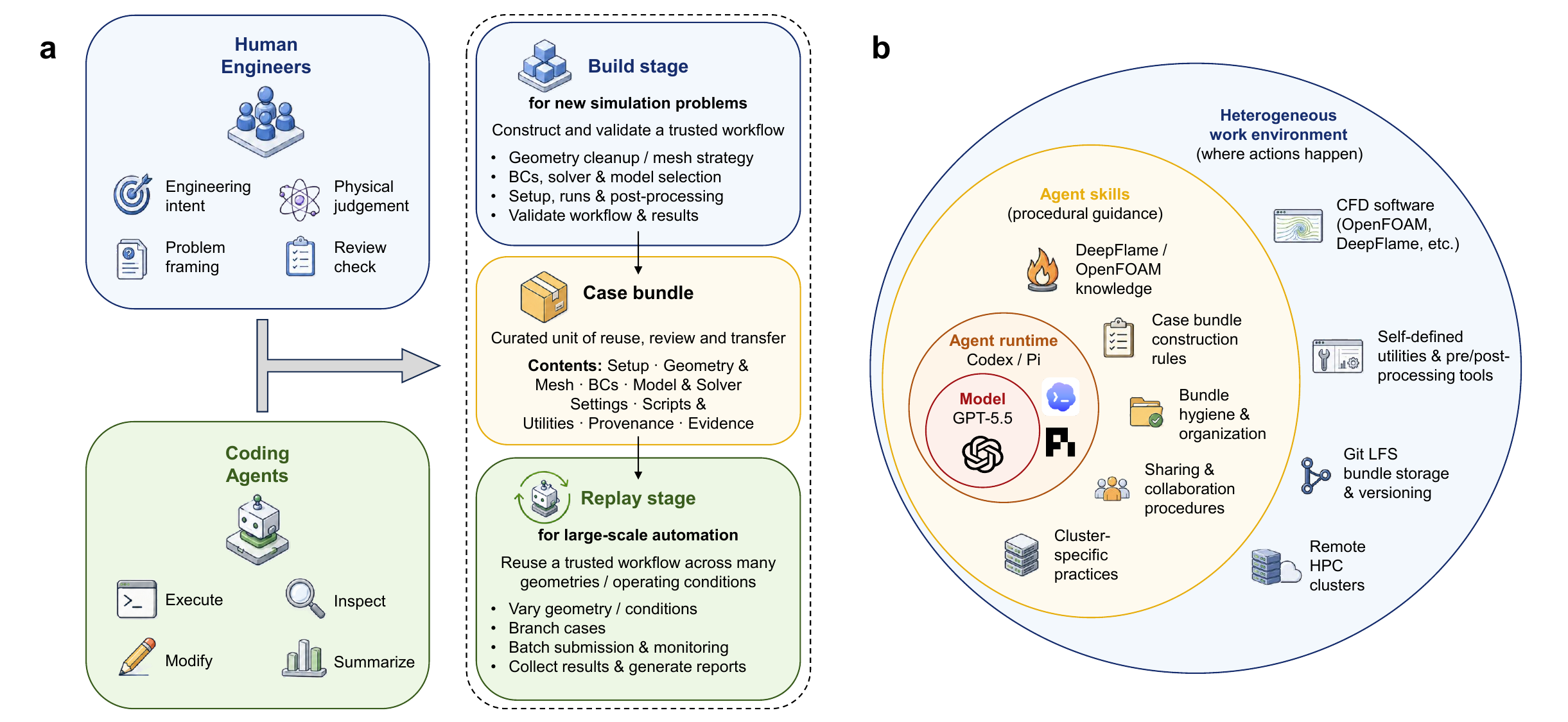}
\caption{How engineers used coding agents in the reported CFD work. Panel (a) shows Build and Replay around reusable case bundles. Panel (b) shows  the agent and its working environment, including Codex/Pi runtimes, agent skills, CFD solvers, local utilities, case storage, and remote execution. }
\label{fig:methods-overview}
\end{figure}

We used coding agents in CFD work through a two-mode process centered on OpenFOAM case bundles. Build is the stage where a case directory is assembled and repaired. Replay is the stage where a reviewed directory is rerun across approved variants. The agent works from the same artifacts CFD engineers already use: solver dictionaries in \texttt{system/}, model data in \texttt{constant/}, initial and boundary fields in \texttt{0/}, geometry files, mesh-generation scripts, run scripts, Slurm submission scripts, solver logs, post-processing scripts, manifests, and README or review notes. Figure~\ref{fig:methods-overview} shows how these modes relate to the case bundle, the agent runtime, and the surrounding engineering environment.

At the start of a case, the agent may need to inspect files and confirm how the directory should be run on the local machine or on the cluster. As the case matures, repeated steps move into scripts, manifests, Slurm job chains, and post-processing wrappers so that later variants can be executed with fewer ad hoc commands. This keeps engineering judgment in the review loop and leaves routine execution to the reviewed procedure.

\subsection{Case bundles and operating modes}
\label{sec:bundle-modes}

A case bundle is the reusable OpenFOAM case directory plus the supporting files that make the calculation understandable and repeatable. In practical terms, it includes the solver dictionaries in \texttt{system/}, the material and model data in \texttt{constant/}, the initial and boundary fields in \texttt{0/}, the STL or other geometry inputs, mesh-generation scripts, run scripts, Slurm submission scripts, solver logs, post-processing scripts, a manifest of important files, and brief README or review notes. Generated \texttt{processor*} directories, scratch files, temporary logs, and large post-processing outputs stay outside the reusable core unless they are needed to explain a correction or document a reported result.

The point of the bundle is to keep the decisions that affect a CFD result in files that can be inspected and rerun. For OpenFOAM work these decisions include geometry interpretation, patch naming and mapping, mesh resolution and quality controls, solver choice, discretization settings, time-step controls, and the definitions used for post-processing. Writing those decisions into scripts, dictionaries, manifests, and notes lets another engineer or agent rerun the case directly from the recorded procedure. In this study, exploratory agent work was converted into reusable scripts after the engineer had checked the mesh path, submission chain, and post-processing procedure.

Build is the case-development mode. The agent works directly in the case directory, staging geometry, editing dictionaries, creating or fixing mesh scripts, checking \texttt{checkMesh}, validating \texttt{setFields}, running pilot solver cases, and adjusting post-processing until the engineer is satisfied that the case is configured correctly. The human stays close to the loop because a run that completes can still use the wrong geometry, wrong patch mapping, or wrong physical assumptions.

Replay is the case-reuse mode. The bundle already contains a reviewed solver setup and execution path, so the agent can apply the same procedure to approved variants, submit batches with Slurm dependencies, monitor logs, collect fields and figures, and merge tables or CSV summaries. Human review still gates the batch before it starts and the interpreted outputs before they are used. If a new geometry, solver instability, or metric discrepancy breaks the approved procedure, the work returns to Build and the bundle is updated.

The boundary between the two modes is whether the case is still being assembled or is already a checked procedure that can be rerun as written. Versioned text files, scripts, manifests, and notes belong in the bundle. Generated \texttt{processor*} directories, transient scratch data, and routine logs stay out unless they are needed as evidence.

\subsection{Model and agent runtimes}
\label{sec:agents}

The main 140-variant study used the GPT-5.5 model \citep{openai_gpt55_system_card_2026} through Codex. We use the term agent runtime to mean the interactive environment in which a model reads files, receives instructions, edits files, runs terminal commands, inspects logs, and reports progress. Codex provided integrated file inspection, patching, terminal-command execution, and progress reporting. The broader working environment included the surrounding case files, software tools, remote computers, and procedural instructions available to the agent.

We conducted a second replay exercise through Pi \citep{zechner_pi_2026}, a lighter-weight runtime that relied more directly on case files, terminal commands, and documentation conventions. Four model backends were used in separate runs, namely GPT-5.4 \citep{openai_gpt54_system_card_2026}, DeepSeek V4 Pro \citep{deepseek_v4_release_2026}, GLM-5.2 \citep{zai_glm52_model_2026}, and Qwen3.7-plus \citep{alibaba_qwen37_model_2026}. Each run received the same reviewed injector bundle, the same STL input, access to SCNet, and the same task of preparing the mesh, configuring and running \texttt{interFoam}, post-processing the calculation, and documenting the procedure.

\subsection{CFD software, utilities, and execution environment}
\label{sec:agent-harness}

The operating model is formulated around files, commands, logs, and outputs shared by multiple solvers. The reported injector study used the OpenFOAM-7 \texttt{interFoam} solver. OpenFOAM-style cases are well suited to this work because they are represented through editable files, command-line utilities, solver logs, and post-processing scripts. We expect the same model to apply most readily to CFD software that exposes inputs, commands, logs, and outputs through comparable file and command interfaces.

The execution environment extended beyond the solver itself. CFD work often requires geometry cleaning, mesh generation, mesh-quality checks, data extraction, plotting, and report assembly. Some utilities came from OpenFOAM, ParaView, VTK, Python, or local scripts, and others were developed during agent-assisted work. This representation brought preprocessing and post-processing into the same file-and-command environment as the case bundle.

We used Git as the version-control layer for solver dictionaries, scripts, manifests, and review notes, with Git Large File Storage (LFS) available for large geometry and mesh files. This structure kept the case bundle reviewable and separated reusable inputs from generated output.

Remote computation formed the outer execution layer. The primary remote platform used in this study was SCNet, a Supercomputing Internet platform \citep{scnet_supercomputing_internet_2026}. The agents accessed SCNet through SSH, worked in remote directories, and used Slurm submission scripts to launch and chain jobs. Job identifiers, scheduler logs, solver logs, and result paths provided the information needed to track execution. This setting tested whether the same bundle and review pattern could extend from local files to remote jobs. Engineers retained responsibility for resource choices and interpretation of completed runs.

\subsection{Discoverability and agent skills}
\label{sec:skills}

Agents had to discover a heterogeneous environment. Some context was available directly from files such as directory names, README files, manifests, solver dictionaries, scripts, logs, and version-control state. Other context had to be supplied as short procedural instruction files, which we call skills. Following the Agent Skills pattern, a skill provides compact discovery metadata and a core instruction file, while detailed reference files and optional scripts are loaded only when relevant \citep{anthropic_agent_skills_2025}. This structure made domain procedures available on demand and kept each task instruction concise.

Two skills were used in the reported work. The CFD skill specified how agents should discover the active software environment, select nearby OpenFOAM examples before case construction, launch new cases cautiously, and document run status. The bundle skill specified how agents should recognize a reusable case bundle, distinguish reusable files from generated output, classify requested changes before editing, and preserve enough provenance for review and replay. Together, these skills made Codex and Pi behavior more consistent across local case directories and remote SCNet runs. Engineers retained responsibility for physical validity and acceptance criteria.

The injector study was checked against case files, Slurm job logs, solver logs, and generated post-processing outputs. We used these materials to confirm case progression, solver completion, post-processing completion, and the integrity of the reviewed bundle. The reported 140 completed variants are the cases for which solver and post-processing outputs were available under the reviewed procedure.

\section{Results}
\label{sec:results}

\subsection{Building and replaying the injector-screening bundle}
\label{sec:injector-build-replay}

The injector-screening case used OpenFOAM-7 \texttt{interFoam} to run VOF simulations of STL design variants. A reviewed case bundle supported 140 completed simulations under a procedure that could be replayed, checked, and repaired. The engineering objective was to establish one geometry-to-simulation procedure and apply it consistently across related injector geometries.

\begin{figure}[!t]
\centering
\includegraphics[width=\textwidth]{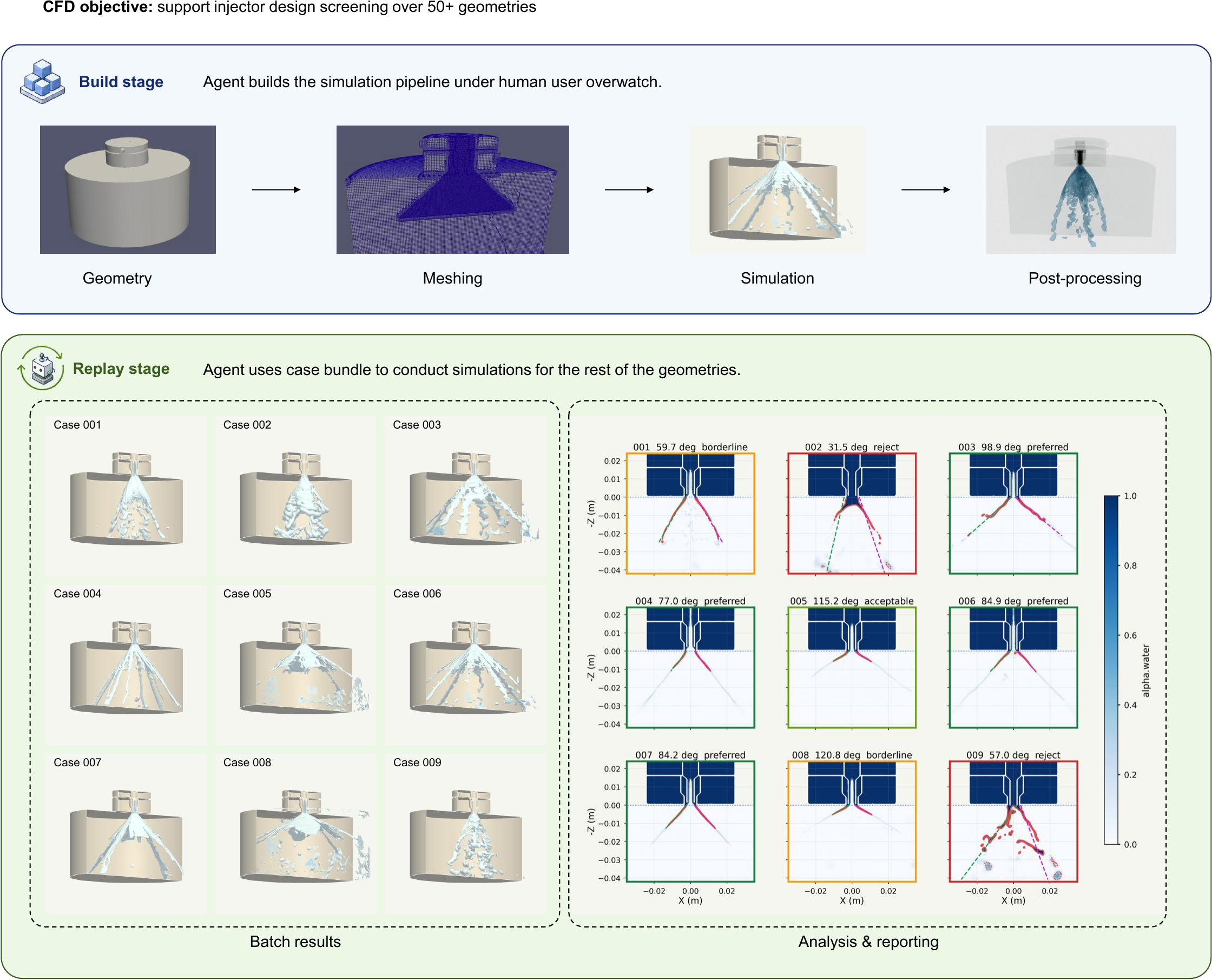}
\caption{Injector screening from Build to Replay. During Build, the agent constructed one OpenFOAM-7 \texttt{interFoam} path from geometry preparation through meshing, simulation, and post-processing under engineering review. During Replay, the reviewed bundle was applied to 140 injector geometries through batch execution. The outputs include center-plane visualizations and cone-angle classifications prepared for engineering inspection.}
\label{fig:injector-case}
\end{figure}

Figure~\ref{fig:injector-case} summarizes the workflow. In Build, the agent and engineer assembled the first working path from STL geometry through meshing, phase initialization, solver execution, and post-processing. In Replay, the reviewed bundle was reused to run the remaining variants, with Slurm providing the batch execution layer and the bundle preserving the case state needed to interpret the outputs.

The reviewed bundle fixed the CFD decisions that mattered for screening. Geometry staging, patch identification, mesh generation, mesh synchronization, \texttt{setFields}, decomposition, and solver dictionaries were all maintained inside the case directory. For this study, the relevant patches were \texttt{fuel}, \texttt{oxy}, \texttt{outlet}, and \texttt{wall\_combustor}. The names \texttt{fuel} and \texttt{oxy} identified the two inlet patches in this non-reacting setup. The solver setup used VOF controls, \texttt{fvSchemes}, \texttt{fvSolution}, gravity, phase properties, and time-control settings appropriate for non-reacting \texttt{interFoam} screening. Before batch execution, the engineer reviewed three initial geometries and checked patch mapping, \texttt{checkMesh}, \texttt{setFields}, decomposition, solver progression, and the resulting post-processing outputs.

After this early verification, the agent consolidated the repeated operations into scripts. For each approved STL geometry, the scripts staged the geometry, generated and synchronized the mesh, initialized the phase field, decomposed the case, submitted dependent Slurm jobs, monitored solver completion, and invoked post-processing. The same procedure was then applied to the remaining 137 variants. Mesh preparation used one compute node with 32 tasks, and the VOF stage used 96 decomposed subdomains across three compute nodes. The target solver time was 0.05~s with output every 0.005~s.

Post-processing produced center-plane visualizations, cone-angle estimates, liquid-film-thickness and breakup-length summaries, per-time records, inlet-patch flow-rate tables, and merged CSV summaries. Center-plane samples were taken from \texttt{alpha.water} and velocity on the x-z plane. Cone angle, liquid-film thickness, and breakup length were computed by the engineer-provided \texttt{postProcess.py} from sampled VTK fields over the available final-time window. Inlet-patch flow rates were extracted at 0.05~s from reconstructed \texttt{phi} using \texttt{surfaceFieldValue}; because the OpenFOAM-7 \texttt{fieldExpression} route was unavailable, the agent used the reconstructed flux path and recorded signed and absolute inlet fluxes for \texttt{fuel} and \texttt{oxy}.

Batch execution exposed several implementation differences that were then absorbed into the scripts. The post-processing wrapper accepted both sampling-output locations encountered in the cases, invoked Python from the relevant case directory, and normalized the resulting center-plane visualizations. Preprocessing was moved from login-shell commands to Slurm jobs after SSH interruptions, and job-state checks were added to prevent overlapping submissions. Each correction was retained in the bundle and applied to subsequent variants.

Replay completed the remaining 137 variants, bringing the full set to 140 completed VOF calculations with post-processing outputs. Completion was checked against Slurm states, solver logs reaching \texttt{Time = 0.05} and \texttt{End}, expected output files, and the first and last case identifiers in the merged summaries. The resulting outputs included center-plane MP4 files, per-case summary rows, per-time records, and inlet-flow records for engineering inspection. These results establish a replayable injector-screening workflow with concrete CFD artifacts and repairable batch execution.

\subsection{Cross-model replay and interaction footprint}
\label{sec:cross-model-replay}

The reviewed bundle was then replayed through Pi with GPT-5.4, DeepSeek V4 Pro, GLM-5.2, and Qwen3.7-plus. Each replay session handled one STL geometry. All four runs succeeded and produced verified CFD results. Each model used the bundle to recover the same working sequence from geometry staging and mesh generation through the \texttt{interFoam} calculation, result handling, and reproducibility documentation. Together, these runs show that a lightweight runtime can use a common procedure across different models.

\begin{figure}[!t]
\centering
\includegraphics[width=\textwidth]{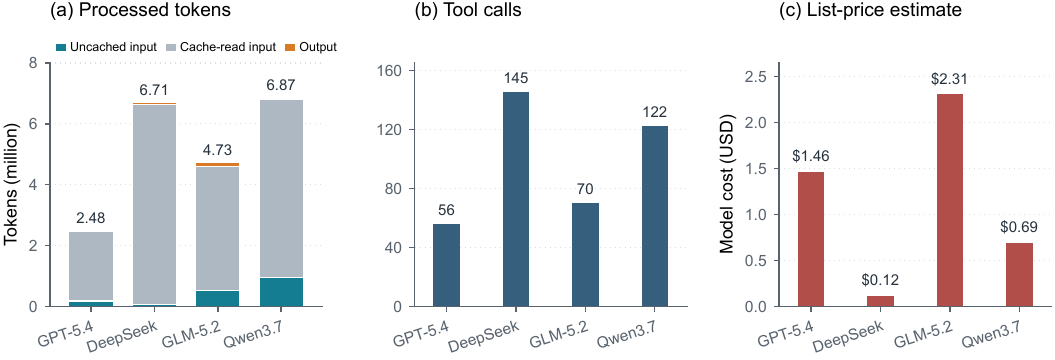}
\caption{Interaction footprint of four successful Pi replays, each applied to one STL geometry. Panel (a) separates uncached input, cache-read input, and output tokens summed over each session. Panel (b) reports tool calls, most of which were shell commands used to inspect files, operate on SCNet, and monitor OpenFOAM jobs. Panel (c) shows API-equivalent list-price estimates in USD based on provider prices.}
\label{fig:cross-model-replay}
\end{figure}

The four runs differed in both tool use and token consumption, as shown in Figure~\ref{fig:cross-model-replay}. GPT-5.4 used 56 tool calls and processed 2.48 million tokens, DeepSeek V4 Pro used 145 tool calls and processed 6.71 million tokens, GLM-5.2 used 70 tool calls and processed 4.73 million tokens, and Qwen3.7-plus used 122 tool calls and processed 6.87 million tokens. Cached input represented more than 85\% of processed tokens in all four sessions, reaching 98\% for DeepSeek V4 Pro. The estimated API costs ranged from USD~0.12 to USD~2.31.

\section{Discussion}
\label{sec:discussion}

The results indicate that the value of coding agents in this workflow depends on how simulation work is organized. The case bundle preserved the state required for reconstruction and reuse, including configurations, geometry processing, meshing, execution, post-processing, and review records. It also provided a location in which corrections could be incorporated into later runs. The successful Pi replays across four LLM models show that this state was usable across different agent configurations.

The two operating modes placed different demands on engineering review. Build involved the formation of the numerical and procedural setup. Geometry interpretation, patch assignment, mesh generation, solver configuration, sampling, and metric definitions were established during this stage. Each of these decisions could affect the meaning of the resulting simulation, so agent-generated files and procedures required inspection before broader use.

Replay operated on a narrower and more stable problem. The agent applied the reviewed procedures to new variants, submitted and monitored Slurm jobs, collected outputs, and executed the associated post-processing. This reduced the number of decisions made during each run and made higher autonomy practical. Review remained necessary for the validity of the geometry, mesh, and reported metrics.

The case therefore supports bounded autonomy through staged delegation. Agents can contribute substantially to case construction and repetitive execution, while engineers retain control over the decisions that establish physical meaning and determine whether the results are suitable for use.

\section{Conclusion}
\label{sec:conclusion}

This study defined a CFD-agent workspace around OpenFOAM case bundles. The bundle brought solver dictionaries, geometry and mesh procedures, run scripts, Slurm submission files, solver and scheduler logs, post-processing code, manifests, and review notes into a persistent working context. These artifacts allowed the case to be inspected, corrected, and reused as the simulation developed. The bundle therefore provided a practical means of preserving the engineering state that coding agents need when working across multiple stages of a CFD calculation.

We applied this arrangement to an OpenFOAM-7 \texttt{interFoam} injector-screening study involving complex STL geometry, custom preprocessing, remote working directories, Slurm execution, and bundled post-processing. After the initial case had been built and reviewed, the procedure was replayed across 140 design variants with solver and post-processing completion. Four additional Pi replays used the same reviewed bundle with different model backends. All four runs succeeded, while their tool use, token consumption, and estimated model costs varied. Together, these results show that a reviewed bundle can preserve a common computational procedure across repeated variants and agent configurations.

The study also established a bounded operating pattern with two modes. Build supports agent-assisted case construction and repair while engineers review geometry interpretation, patch mapping, mesh suitability, solver settings, remote execution assumptions, and post-processing definitions. Replay applies the reviewed procedure to approved variants and allows the agent to perform more of the repetitive work, including job submission, monitoring, result collection, and batch analysis. Engineers retain responsibility for the physical meaning and suitability of the resulting calculations.

These findings suggest several design implications for future CFD-agent environments. Agent work benefits from structured case directories, persistent case state, executable scripts, recorded outputs, and reviewable decision points. As exploratory work is converted into tested procedures, scripting can reduce repeated model-mediated actions and extend the range of variants that can be processed consistently. This provides a practical basis for bounded autonomy in CFD while keeping engineering judgment at the points where it is needed most.

\bibliographystyle{cnf-num}
\bibliography{references}

\end{document}